# Effective *g*-factors of Carriers in Inverted InAs/GaSb Bilayers

Xiaoyang Mu[1,2], Gerard Sullivan[3], and Rui-Rui Du[1,2,4,]

[1]*International Center for Quantum Materials, School of Physics,*
*Peking University, 100871, Beijing, China*
[2]*Collaborative Innovation Center of Quantum Matter, 100871, Beijing, China*
[3]*Teledyne Scientific and Imaging, Thousand Oaks, CA 91630, USA*
[4]*Department of Physics and Astronomy, Rice University, Houston, Texas 77251-1892, USA*

## *Abstract*

We perform tilt-field transport experiment on inverted InAs/GaSb which hosts quantum spin Hall insulator. By means of coincidence method, Landau level (LL) spectra of electron and hole carriers are systematically studied at different carrier densities tuned by gate voltages. When Fermi level stays in the conduction band, we observe LL crossing and anti-crossing behaviors at odd and even filling factors respectively, with a corresponding *g*-factor of 11.5. It remains nearly constant for varying filling factors and electron densities. On the contrary, for GaSb holes only a small Zeeman splitting is observed even at large tilt angles, indicating a *g*-factor of less than 3.

Two-dimensional topological insulator, also known as quantum spin Hall insulator (QSHI), is of great interest for its helical edge states transport properties, which are considered promising for realizing electrical-control of spin transport.[1] It is also proposed theoretically that QSHI can be utilized as an unique building block for proximity-induced unconventional superconductors.[2] The helical edge states are a pair of time-reversal symmetry (TRS) protected counter-propagating one-dimensional modes with spin-momentum locking properties, persisting along the sample boundary. Within single-particle picture, the helical edge states should not experience backscattering in the presence of non-magnetic impurities. On the other hand, it is anticipated theoretically that an external magnetic field breaks TRS, manifesting a decreased edge conductance, for example, as is shown in HgTe/CdTe quantum wells.[3] This property has recently been examined in the QSHI made of inverted InAs/GaSb bilayers where quantized conductance plateau has been observed in micrometer size Hall bar devices. Unexpectedly, in InAs/GaSb bilayer, the quantized conductance keeps constant for an in-plane magnetic field as large as 12T.[4] It would be important to know the Zeeman energy scale in this system before any reasonable interpretation can be made. It is well known that valence *g*-factor in heterostructures is strongly anisotropic, and its in-plane component is often close to zero.[5] It is of particular interest to determine the effective *g*-factor in InAs/GaSb bilayers, where conduction and valence bands are hybridized. Note the value of *g*-factor of this system is referring to the bulk carriers, but this study should also help to understand the Zeeman energy scale in the edge spectrum.

The behaviors of Landau levels and Zeeman splitting under an external magnetic field are explicitly understood in a single-particle frame. Generally, when the magnetic field is oriented perpendicular to a two-dimensional electron gas (2DEG), Zeeman splitting energy is much smaller than the Landau level splitting. Nevertheless, it is known that cyclotron energy $E_C$ is proportional to the perpendicular magnetic field $B_\perp$, while Zeeman energy $E_Z$ is proportional to the total magnetic field $B_{tol}$. Thus by rotating the magnetic field, $E_Z$ becomes comparable to $E_C$ at certain tilt angle, known as coincidence conditions [6, 7]. This method has been utilized to determine the effective *g*-factor in various materials with considerations of non-interacting electrons [6, 8, 9] or with exchange enhancement [7, 10, 11, 12]. In the single-particle picture, coincidence conditions are characterized by the parameter *r*, the ratio of Zeeman and cyclotron energy, $r = g\mu_B B_{tol}/\hbar\omega_C$.

In this paper, we report a systematic study of the coincidence spectra in InAs/GaSb bilayer system by a tilted magnetic field. Remarkably, we observe anti-crossing behaviors at even integer filling factors and regular crossings at odd filling factors respectively for Fermi levels staying in the conduction band, giving a *g*-factor of ~11.5. We further conclude that *g*-factor of InAs electrons keeps roughly constant for various magnetic fields, densities and different crystalline orientations. Moreover, a small *g* value of less than 3 is deduced for GaSb holes. Possible origins of the anti-crossings and the issue regarding *g*-factor will be discussed.

The samples used in this experiment consist of an 11nm wide InAs and 7nm GaSb quantum well embedded in two 50nm $Al_{0.8}Ga_{0.2}Sb$ barriers, as shown in Fig.1a. When the system is in inverted regime (InAs conduction band falls below GaSb valance band), electron and hole carriers hybridize by tunneling process, opening up a mini gap [13] (Fig.1b). Use of GaSb substrate enables a high mobility of $\mu$=40,000cm$^2$/Vs at a relatively low electron density $n$=2.3×10$^{11}$cm$^{-2}$. The samples are patterned into two Hall bars of the same size (75×25um), one of which is at 45 degree angle to the other on the same chip [marked by Sample A, aligned with (001); And Sample B, aligned with (110) respectively]. This design is for probing the samples' anisotropic properties. Ohmic contacts are made by indium with annealing process. An aluminum Schottky gate is used as front gate ($V_{front}$) for tuning across the topological regime (Fig.1c). Here we note that the distinct peak values are sensitive to different cooling cycles as a result of residual bulk carriers and anisotropy of carrier scattering length. The samples are mounted on a revolving sample stage in a He-3 cryostat with a base temperature of 300mK and a magnetic field of up to 15T. Transport measurements are performed using standard lock-in techniques with an excitation current of $I$=100 nA and frequency $f$=17 Hz.

Firstly, we determine the effective mass $m^*$(InAs) and $m^*$(GaSb), respectively, in this hybridized bilayer system from temperature dependent Shubnikov-de Haas (SdH) oscillation analysis. At low magnetic field, the small-amplitude SdH oscillations in longitudinal resistance $R_{xx}$ are generally described by

$$\Delta R_{xx} = 4R_0 \frac{X_T}{\sinh(X_T)} exp\left(-\frac{\pi}{\omega_c \tau_q}\right) cos\left(\frac{2\pi E_F}{\hbar \omega_c} - \pi\right) cos\left(\pi \frac{E_z}{\hbar \omega_c}\right)$$

Where $R_0$ is the zero-field resistivity, the term $X_T = 2\pi^2 k_B T/\hbar \omega_c$ describes thermal damping with $k_B$ being the Boltzmann constant, $\tau_q$ is the quantum lifetime charactering disorder, and $E_F$ is the Fermi energy.[12, 14, 15, 16] The effective mass $m^*$ could thus be extracted from the SdH amplitude $A(T)$ by fitting to $\ln(A/T)$ versus $T$ plot. Fig.1d displays temperature dependence data of Sample A at zero gate bias, and Fig.1e gives an example of linear fitting to the SdH amplitude, indicating an electron effective mass $m^*$(InAs) = 0.040$m_0$, where $m_0$ is the free electron mass. Moreover, $m^*$(InAs) increases slightly with the magnetic field as a result of band non-parabolicity, and keeps nearly invariable for various densities ($n$=3.42×10$^{11}$cm$^{-2}$, $n$=2.38×10$^{11}$cm$^{-2}$ and $n$=1.38×10$^{11}$cm$^{-2}$) shown in Fig.1f. This result is consistent with previously reported value around 3T,[8] and the intercept at $B$=0 gives an effective mass of 0.032$m_0$, which is in excellent agreement with theoretical value.[17] On the other hand, $m^*$(InAs) was reported to be nearly constant with tilt angles.[9] Thus the fitting $m^*(\text{InAs}) = (0.032 + 0.005 \times B)m_0$ from Fig.1f will be used for coincidence analysis. The quantum life time $\tau_q$ = 0.24ps is extracted from Dingle analysis by fitting $\ln(\Delta R_{xx}\sinh(X_T)/4R_0X_T)$ to $1/B$.[18] Similarly, from the temperature dependence of SdH amplitude at $V_{\text{front}}$= -0.475V (Fig.1g), we obtain hole effective mass $m^*$(GaSb) = $0.136m_0$, which is smaller than commonly agreed value of $0.3m_0$.[17] This could originate from some hybridization of electron and hole carriers.

Fig.2a displays the longitudinal magneto-resistance as a function of perpendicular magnetic field for various tilt angles at zero gate bias (Sample A). The rotation angle $\theta$, between total magnetic field and sample normal, is accurately determined from Hall resistance $R_{xy}$, and the curves are shifted vertically proportional to $1/cos\theta$ for clarity. Remarkably, crossings and anti-crossings of LLs are observed at even and odd filling factors (red lines to guide the eyes). We focus on the crossing at $v$=5 where an initial Zeeman splitting occurs at small $\theta$ with two separate peaks, followed by a mergence into a single peak at ~82.5º. This crossing corresponds to the ratio $r$ = $E_Z/E_C$ =2 as illustrated in the sketch of energy spectrum (See inset of Fig.2a, marked by red dot). At around $v$=4 where single particle picture predicts a crossing of spin split LLs, SdH minima weaken with satellite peaks coming closer to $\theta$~76.2º, and enhance again before formation of a single peak. This absence of gap closing and associated anti-crossing behavior suggest that many-body effects or spin-mixing terms may present in this system, so that the spin-resolved Landau levels are strongly modified (Fig.2a inset, marked by red arcs). We now analyze the effective $g$-factor in this regime. From the coincidence conditions, $g$-factor is determined by g = $r*2m_0 cos\theta/m^*$, and the systematic error is relatively large at large tilt angles. To determine the precise position of the coincidences, the changes of resistance extrema at integer filling factors as a function of $1/cos\theta$ are plotted in fig.2b. As expected an $R_{xx}$ maximum (minimum) value is observed for even (odd) filling factor at r=1. The extreme position $1/cos\theta$=4.4, 4.1, 4.6 and corresponding $g$-factor 11.3, 12.2, 10.9 are obtained for filling factors $v$=4, 5, 6 respectively, as shown in fig.2c. Thus the estimated g-factor of electrons is ~11.5. In addition, using the coincidence position at ν=5, the crossing angle at r=2 is 83°, which is a little beyond experimental accessible magnetic field and close to our estimation from fig.2a.

We further study LL spectra at various electron densities in Fig.3, namely, for $n_1$=1.34×10$^{11}$cm$^{-2}$ (at $V_{front}$ = -0.1V), $n_2$=3.34×10$^{11}$cm$^{-2}$ (at $V_{front}$ = 0.1V) and $n_3$=3.90×10$^{11}$cm$^{-2}$ (at $V_{front}$ = 0.15V). The LL crossing and anti-crossing behaviors also emerge at roughly similar cross angles (Fig.3a-c). Detailed analyses of the $R_{xx}$ extrema with respect to rotation are given in Fig.3d-f. Taking $V_{front}$=0.15V as an example, in fig.3f, the coincidence positions for filling factor ν=7~11 are $1/cos\theta$=4.2, 4.6, 4.2, 4.8, 4.0, and related g-factors are 11.8, 10.9, 11.8, 10.4, 12.4 respectively. Thus the electron g-factor keeps nearly unchanged for different magnetic fields within small derivations. We then summarize all the averaged g values as a function of electron densities in fig.4a, where the upper and lower limits of g-factors for different magnetic field at certain density are indicated with error bars. We further deduce that electron g-factor remains constant for density changes.

For GaSb hole carriers, only small Zeeman splitting is observed at very large tilt angle $\theta$~79° shown in Fig.4b. Thus the related g-factor is within the range $g < 2m_0 cos\theta/m^*(GaSb) \approx 3$. Since hole g-factor is small, a rough estimation could be made from LL broadening $\Gamma$ by comparing the onset of SdH oscillations (at $B_0$=0.7T) with the Zeeman energy where spin split could be resolved. We thus have $\Gamma \approx \hbar eB_0/m^* \approx g\mu_B B_{tol}$, giving g ~0.9 ($B_{tol}$=11.2T at 77°).

In the last part, we apply the same analyses to Sample B with the current aligned with the (110) crystalline direction. Temperature dependent SdH measurements are also performed giving an electron effective mass of $m^*(InAs) = (0.031 + 0.006 \times B)m_0$. Crossings and anti-crossings of LLs are also reviewed at odd and even filling factors respectively, for various electron densities we have studied ($n_1$=1.39×10$^{11}$cm$^{-2}$, $n_2$=2.38×10$^{11}$cm$^{-2}$ and $n_3$=3.96×10$^{11}$cm$^{-2}$). Fig.4c displays an example of the resistance change with $1/cos\theta$ at density $n_2$=2.38×10$^{11}$cm$^{-2}$ ($V_{front}$ = 0V), giving g-factors of 10.5, 11.4, 11.7 for filling factor $v$=4~6. Fig.4d summarizes the electron g-factors for Sample B versus densities, and we obtain $g$~11.5. These results strongly suggest that the g-factor of electron type is isotropic, and it nearly stays constant for a large range of gate bias. As for hole type carriers in this sample, we didn't acquire clear SdH oscillations, because the longitudinal resistance keeps around 20kΩ for much negative front gate bias and shows low mobility. This could result from the anisotropic property of valence band in InAs/GaSb bilayer system.[19]

We now discuss the possible origin of these observed LL anti-crossings. This nontrivial behavior has been previously reported in several materials, i.e. in Ga$_x$In$_{1-x}$As/InP heterostructure[7], in InAs/AlSb quantum well[9], and in InAs/InGaAs/InAlAs quantum well[20]. Many-body interactions and spin mixing terms are considered, such as electron-electron interaction, exchange interaction and spin-orbital (SO) interaction. Giuliani and Quinn predicted a first-order transition from a spin-unpolarized state to spin-polarized state at filling factor $v$=2 when taking into account the electron-electron interaction[21], and this was experimentally observed in Ref [7] by non-vanishing QH minima. However, this transition could only occur at small filling factors, where the magnetic length $l = \sqrt{\hbar/eB}$ is small and Coulomb interaction $e^2/l$ is large. In our experiment, anti-crossing at $v$=10 and $B$=1.6T is observed, thus electron-electron interactions should not be our scenario. Secondly, in the presence of exchange interactions, one would expect a magnetic field dependence of the effective g-factor near the onset of spin-splitting[11, 12, 20]. We currently observe roughly constant g-factor behavior for low magnetic fields, and the effect of exchange enhancement may not be adequately explored until high magnetic field experiments are performed.

Finally, SO interaction can play an important role but could only couple certain pair of Landau levels. The level mixings at $r$=1 and $r$=3... are allowed and can lead to anti-crossings at even filling factors, whereas $r$=2 is forbidden due to selection rules[20, 22, 23]. This is very close to our case. We also note that InAs-based materials have a large Rashba SO interaction, and experimentally we find a resistance dip around zero field known as weak anti-localization, which can be understood as a type of weak localization by including Rashba effects.

Next we turn to discuss the change of $g$-factor in InAs/GaSb bilayer system. As we have previously demonstrated, the electron $g$-factor is as large as 11.5, and shows no signature of decreasing with density. Even in the hole type region, there is still a finite $g$-factor of around 0.9, yielding a Zeeman energy of 4meV at 12T. This value is larger than the localization gap (26K) reported in Ref [4], and is large enough to induce partial spin-polarization. Moreover, the isotropic property of $g$-factor means that this energy scale is applied equally parallel or perpendicular to the edge. So the single particle parameters, such as bulk $g$-factor should not be attributed to the issue why InAs/GaSb QSH plateau shows no gap closing with large in-plane magnetic fields. On the other hand, this system is in a strongly interacting regime, for example, manifesting Luttinger liquid in the edge states.[24] The robust edge transport in InAs/GaSb bilayers as reported in Ref [4] may not be adequately explained until the interesting many-body correlations in this system are fully explored.

In conclusion, we have experimentally studied the Landau level spectra in InAs/GaSb bilayers at various front gate biases. LL crossing and anti-crossing behaviors are repeatedly observed at odd and even filling factors, giving an electron $g$-factor of 11.5. It remains nearly constant for various magnetic fields, densities and crystalline orientations. We associate this anti-crossing behavior with strong spin-orbital interactions. For hole type carries, only small Zeeman splitting is seen at large tilt angle, giving a $g$-factor of less than 3.

**Acknowledgements**

We thank Li Lu and Changli Yang for discussions. Use of equipment in IOP, CAS is gratefully acknowledged. The work at Peking University was supported by NBRPC Grants (No. 2012CB921301 and No. 2014CB920901). R. R. D. was supported by NSF Grants (No. DMR-1207562 and No. DMR-1508644), and Welch Foundation Grant (No. C-1682).

**References**

[1] M. Z. Hasan and C. L. Kane, Rev. Mod. Phys. **82**, 3045 (2010).

[2] X. L. Qi and S. C. Zhang, Rev. Mod. Phys. **83**, 1057 (2011).

[3] M. Konig, S. Wiedmann, C. Brune, A. Roth, H. Buhmann, L. W. Molenkamp, X. L. Qi and S. C. Zhang, Science **318**, 766 (2007).

[4] L. J. Du, I. Knez, G. Sullivan and R. R. Du, Phys. Rev. Lett. **114**, 096802 (2015).

[5] S. Y. Lin, P. Wei and D. C. Tsui, Phys. Rev. B **43**, 12110 (1991).

[6] F. F. Fang and P. J. Stiles, Phys. Rev. **174**, 823 (1968).

[7] S. Koch, R. J. Haug, K.V. Klizing and M. Razeghi, Phys. Rev. B **47**, 4048 (1993).

[8] T. P. Smith III and F. F. Fang, Phys. Rev. B **35**, 7729 (1987).

[9] S. Brosig, K. Ensslin, A. G. Jansen, C. Nguyen, B. Brar, M. Thomas and H. Kroemer, Phys. Rev. B **61**, 13 045 (2000).

[10] R. J. Nicholas, R. J. Haug, K. V. Klitzing, and G. Weimann, Phys. Rev. B **37**, 1294 (1988).

[11] W. Desrat, F. Giazotto, V. Pellegrini, F. Beltram, F. Capotondi, G. Biasiol and L. Sorba,

Phys. Rev. B, **69**, 245324 (2004).

[12] X. J. Wu, T. X. Li., C. Zhang and R. R. Du, Appl. Phys. Lett. **106**, 012106 (2015).

[13] M. J. Yang, C. H. Yang, B. R. Bennett, and B. V. Shanabrook, Phys. Rev. Lett. **78**, 4613 (1997).

[14] A. Isihara and L. Smrcka, J. Phys. C: Solid State Phys. **19**, 6777 (1986).

[15] S. A. Tarasenko, Phys. Solid State, **44**, 1769 (2002).

[16] P. T. Coleridge, Phys. Rev. B, **44**, 3793 (1991).

[17] N. Bouarissa and H. Aourag, Infrared Phys. & Tech. **40**, 343 (1999).

[18] B. Das and S. Subramaniam, M. R. Melloch and D. C. Miller, Phys. Rev. B **47**, 9650 (1993).

[19] M. Lakrimi, S. Khym, R. J. Nicholas, D. M. Symons, F. M. Peeters, N. J. Mason, and P. J. Walker, Phys. Rev. Lett **79**, 3034 (1997).

[20] W. Desrat, F. Giazotto, V. Pellegrini, M. Governale and F. Beltram, F. Capotondi, G. Biasiol and L. Sorba, Phys. Rev. B **71**, 153314 (2005).

[21] G. F. Giuliani and J. J. Quinn, Phys. Rev. B **31**, 6228 (1985).

[22] B. Das, S. Datta, and R. Reifenberger, Phys. Rev. B **41**, 8278 (1990).

[23] J. C. Chokomakoua, N. Goel, S. J. Chung, M. B. Santos, J. L. Hicks, M. B. Johnson, and S. Q. Murphy, Phys. Rev. B **69**, 235315 (2004).

[24] T. X. Li, P. J. Wang, H. L. Fu, L. J. Du, K. A. Schreiber, X. Y. Mu, X. X. Liu, G. Sullivan, G. A. Csáthy, X. Lin and R. R. Du, Phys. Rev. Lett **115**, 136804 (2015).

# Figure Captions

**FIG. 1.** (Color online) (a) Detailed information of wafer structure. Red and green arrows mark the electron and hole channels respectively. (b) Band structure of inverted regime. Two bands hybridize and open up a gap $\Delta$. (c) Longitudinal resistance $R_{xx}$ as a function of front gate voltage ($V_{front}$) at 300mK for two Hall bar (75×25um) on the same chip. Sample B (red curve) is at 45 degree angle to Sample A (blue curve) for studying anisotropy effect. (d) Temperature dependence of magneto-resistance at zero front gate bias of Sample A. The SdH amplitudes $A$ divided by temperature $T$ are linearly fitted to $T$ in (e), giving an electron effective mass of $m^*(InAs)=0.040m_0$. (f) InAs effective mass at various densities (open triangles for $n=3.42\times10^{11}cm^{-2}$, squares for $n=2.38\times10^{11}cm^{-2}$ and open circles for $n=1.38\times10^{11}cm^{-2}$). (g) Temperature dependence of SdH oscillations at $V_{front}=-0.475V$ (hole type). The hole effective mass $m^*(GaSb)=0.136m_0$ is obtained.

**FIG. 2.** (Color online) (a) Longitudinal resistance $R_{xx}$ as a function of the perpendicular magnetic field for different tilt angles measured at 300mK for Sample A ($V_{front}=0V$). Curves are shifted proportional to $1/cos\theta$ for clarity, where $\theta$ is the angle of total magnetic field with respect to sample normal (left inset). Vertical thin dotted lines indicate integer filling factors. Crossings of LLs at $v=5$ are marked by red dashed lines. Oscillation peak positions are joint by red arcs. Right inset: schematic illustration of LL crossings and anti-crossings, where spin-split LLs vs $1/cos\theta$ is plotted. The positions where LL anti-crossing is observed are marked by red arcs, and the crossing point of $v=5$ are indicated by red dot. (b) $R_{xx}$ extrema for filling factors $v=4, 5, 6$ as a function of $1/cos\theta$. Curve for v=6 is vertically shifted up by 0.5kΩ for clarity. The arrows show the values used to calculate *g*-factors. (c) Calculated *g*-factors from (b) at different magnetic fields.

**FIG. 3.** (Color online) (a)-(c) More crossings of Landau levels under titled magnetic field at different electron densities (a) $n=1.34\times10^{11}$cm$^{-2}$ ($V_{\text{front}} = -0.1$V), (b) $n=3.34\times10^{11}$cm$^{-2}$ ($V_{\text{front}} = 0.1$V) and (c) $n=3.90\times10^{11}$cm$^{-2}$ ($V_{\text{front}} = 0.15$V). The related $R_{xx}$ extrema are plotted as a function of $1/cos\theta$ at certain integer filling factors: (d) $v=3, 4$ for $V_{\text{front}} = -0.1$V, (e) $v=7, 8, 9$ for $V_{\text{front}} = 0.1$V, and (f) $v=7\sim11$ for $V_{\text{front}} = 0.15$V.

**FIG. 4.** (Color online) (a) Values of electron *g*-factors versus densities. The derivations from averaged value caused by magnetic field difference at a certain density are included in the error bar. (b) Magneto-resistance of hole carriers as a function of perpendicular magnetic field at different tilt angles. Only Zeeman splitting is found at large tilt angle indicating small *g*-factor of GaSb. (c) $R_{xx}$ extrema at filling factors $v=4, 5, 6$ as a function of $1/cos\theta$ for Sample B at zero gate bias ($n=1.34\times10^{11}$cm$^{-2}$). (d) Values of electron *g*-factors versus densities for Sample B.

**Figures:**

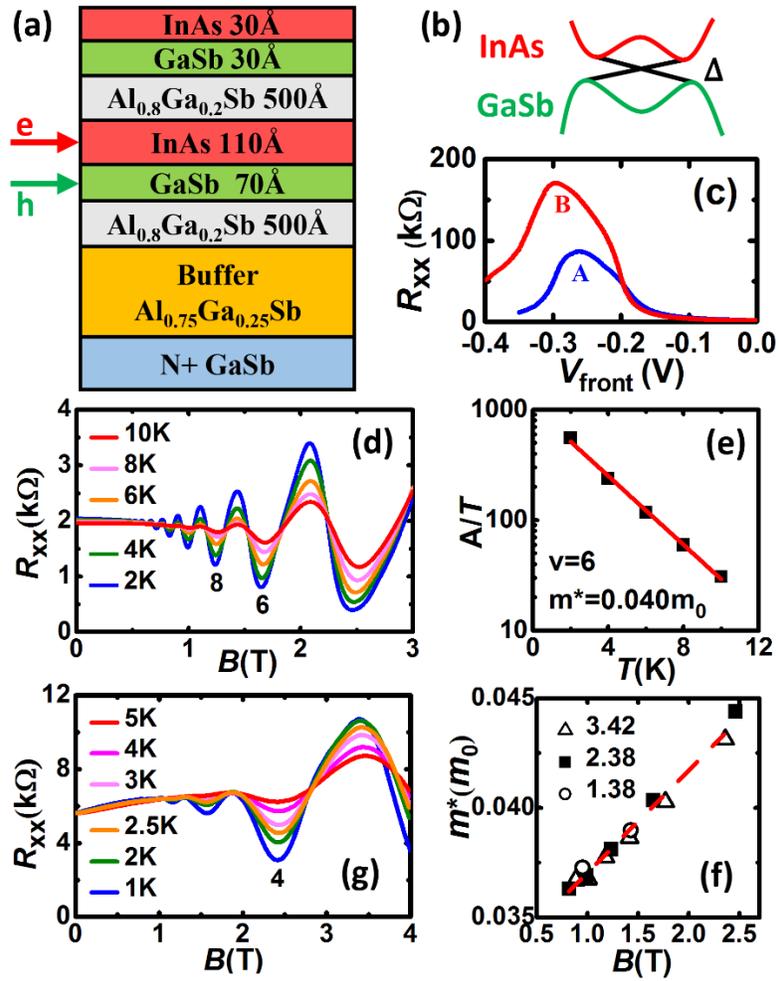

**Figure 1**

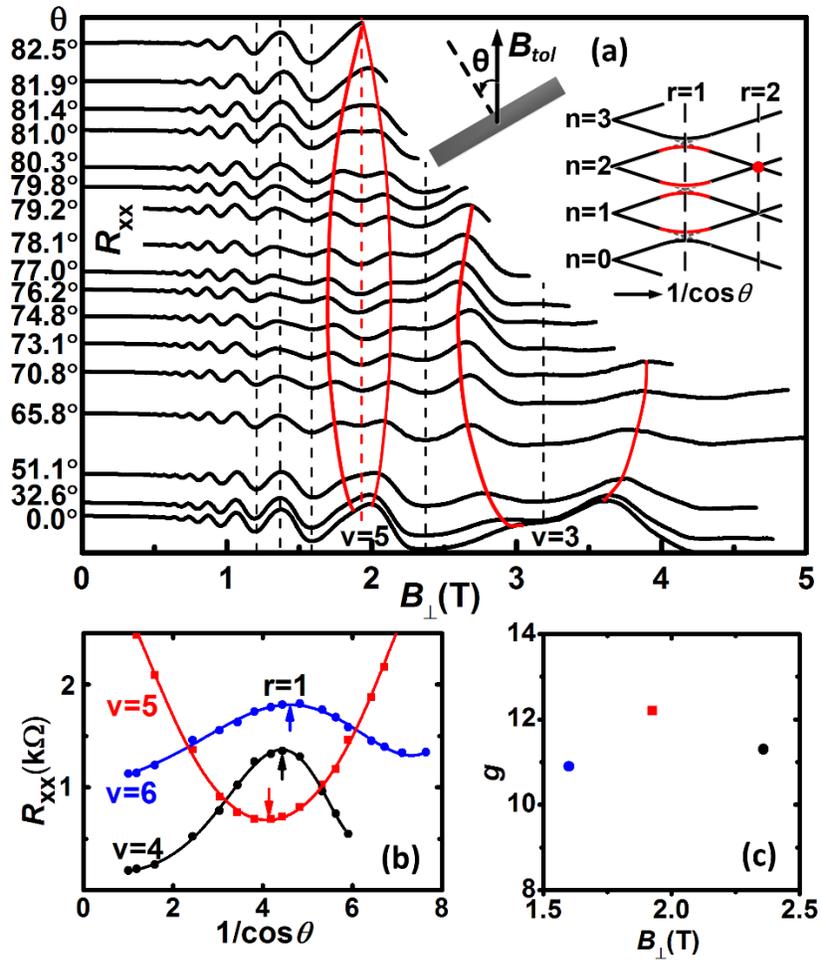

**Figure 2**

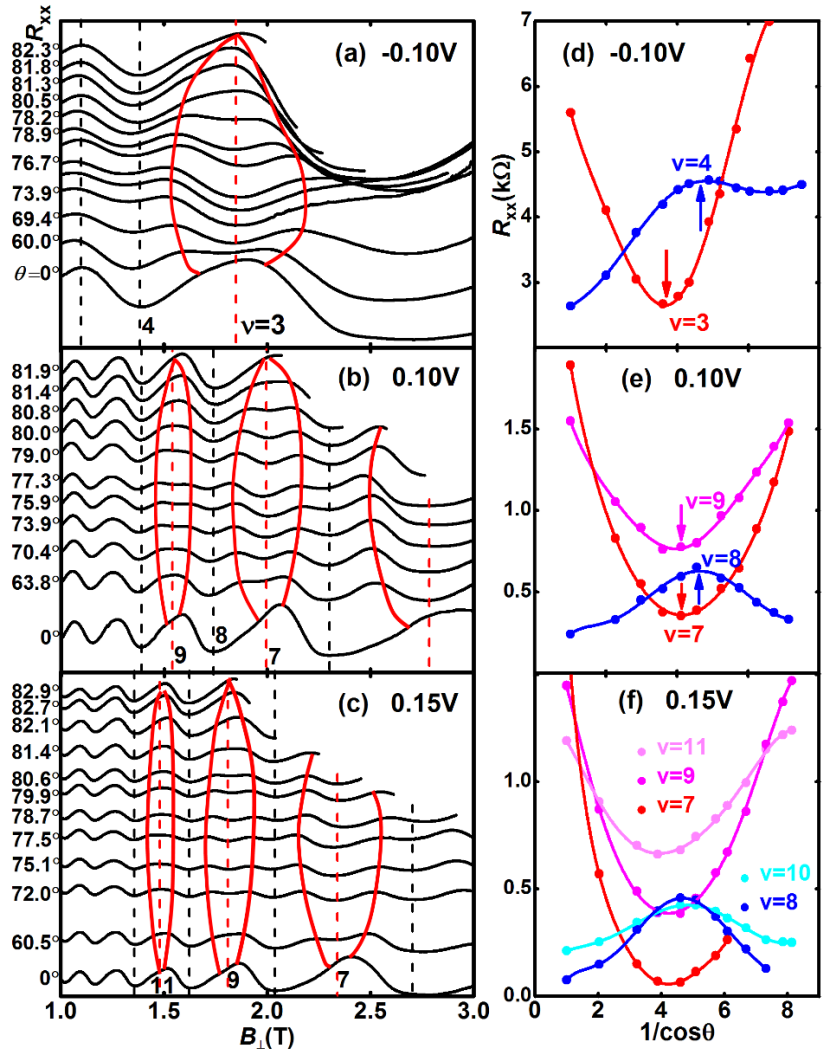

**Figure 3**

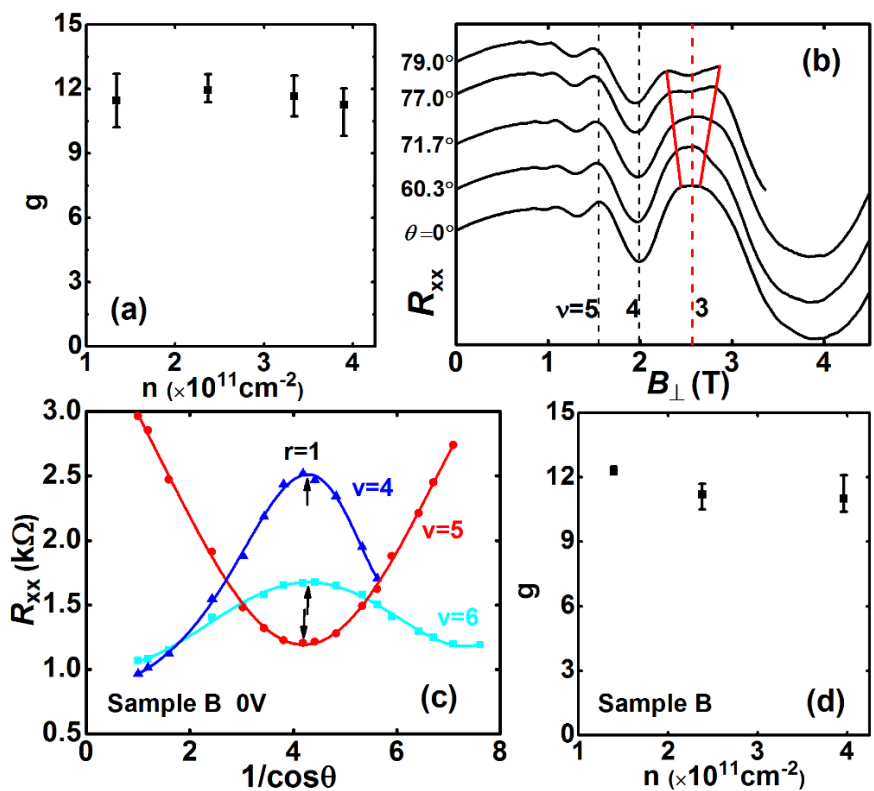

**Figure 4**